# The Infrared Imaging Spectrograph (IRIS) for TMT: Volume Phase Holographic Grating Performance Testing and Discussion

Shaojie Chen*[a], Elliot Meyer[a,b], Shelley A. Wright[a,b], Anna M. Moore[c], James E. Larkin[d], Jerome Maire[a], Etsuko Mieda[a,b], and Luc Simard[e]

[a] Dunlap Institute for Astronomy & Astrophysics, University of Toronto, ON, Canada, M5S 3H4
[b] Department of Astronomy & Astrophysics, University of Toronto, ON, Canada, M5S 3H4
[c] Caltech Optical Observatories, California Institute of Technology, Pasadena, CA, USA, 91125
[d] Department of Physics of Astronomy, University of California, Los Angeles, CA, USA, 90095
[e] Dominion Astrophysical Observatory, National Research Council, Victoria, BC, Canada, V9E 2E7

**ABSTRACT**

Maximizing the grating efficiency is a key goal for the first light instrument IRIS (Infrared Imaging Spectrograph) currently being designed to sample the diffraction limit of the TMT (Thirty Meter Telescope). Volume Phase Holographic (VPH) gratings have been shown to offer extremely high efficiencies that approach 100% for high line frequencies (i.e., 600 to 6000l/mm), which has been applicable for astronomical optical spectrographs. However, VPH gratings have been less exploited in the near-infrared, particularly for gratings that have lower line frequencies. Given their potential to offer high throughputs and low scattered light, VPH gratings are being explored for IRIS as a potential dispersing element in the spectrograph. Our team has procured near-infrared gratings from two separate vendors. We have two gratings with the specifications needed for IRIS current design: 1.51-1.82μm (H-band) to produce a spectral resolution of 4000 and 1.19-1.37μm (J-band) to produce a spectral resolution of 8000. The center wavelengths for each grating are 1.629μm and 1.27μm, and the groove densities are 177l/mm and 440l/mm for H-band R=4000 and J-band R=8000, respectively. We directly measure the efficiencies in the lab and find that the peak efficiencies of these two types of gratings are quite good with a peak efficiency of ~88% at the Bragg angle in both TM and TE modes at H-band, and 90.23% in TM mode, 79.91% in TE mode at J-band for the best vendor. We determine the drop in efficiency off the Bragg angle, with a 20-23% decrease in efficiency at H-band when 2.5° deviation from the Bragg angle, and 25%-28% decrease at J-band when 5° deviation from the Bragg angle.

## 1 INTRODUCTION

The Infrared Imaging Spectrograph (IRIS)[1] is the first light diffraction-limited instrument designed for the future Thirty Meter Telescope (TMT). Working with the advanced adaptive optics system (NFIRAOS)[2,3] and integrated on-instrument wavefront sensors (OIWFS)[4], IRIS is designed to sample the diffraction limit. IRIS combines a powerful near-infrared imager[5] and an integral field spectrograph[6] operating over the ground-based near-infrared band passes with a range from 0.84μm to 2.4μm. A spectral resolution of R=4000 will be used for all broad bands, and a higher resolution mode (R=8000) will potentially be provided for certain modes. Both the imager and spectrograph are designed around 4K by 4K HgCdTe detectors from Teledyne (Hawaii 4RG)[7]. Many of the TMT and IRIS science cases require high sensitivity limits (Barton et al. 2010)[8] and the IRIS technical team is interested in exploring how to maximize the efficiency. Typically in a spectrograph the largest degradation of total throughput is from the detector quantum efficiency, quality and number of optical elements, and the type and quality of the dispersing element (such as, grating, prism, and other diffraction structures). Surface relief diffraction gratings (like ruled gratings and holographic gratings) are traditional dispersing elements for the majority of astronomical spectrographs. Usually, the peak efficiency of traditional diffraction gratings are of order ~80%. Grating vendors are working to optimize the manufacturing process to get higher efficiencies, and optical engineers continue to invent new types of diffraction structures or innovation in their applications. VPH gratings are a type of transmission grating with high diffraction efficiency with low scattered light properties, which have been investigated since the late 1960s[9,10,11].

*sjchen@dunlap.utoronto.ca; phone: 1-416-978-6779

National Optical Astronomy Observatories (NOAO) measured the performance of VPH gratings for astronomy applications in 1998, and got promising results[12]. In the following years, VPH gratings have been applied in different astronomical spectrographs (e.g. Clemens et al. 2000, Bershady et al. 2008, Hou et al., 2010, Laurent et al. 2010, Zanutta et al. 2014)[13-18]. The introduction of VPH gratings to optical astronomy has brought greater efficiencies, as well as offered more flexibility in the optical design of instruments[19].

Compared with traditional surface relief gratings, VPH gratings have some unique features, such as high transitive efficiency in the primary operating mode (theoretical diffraction efficiencies can approach 100% for high line frequencies, 600 to 6000l/mm)[20] and environmentally stable structure with encapsulation between two glass substrates, which are applicable for astronomical spectrographs. VPH gratings have been shown to be viable in optical astronomical instrumentation (SOAR, WIYN, LOMST, MUSE, and AFOSC, FOCAS)[13-18], but have been less exploited at near-infrared wavelengths (Arns et al. 2010 for APOGEE)[21].

In this paper, we investigated two types of VPH gratings designed for IRIS to assess the quality and the feasibility of these gratings for its spectrograph. The first grating is designed for H-band with a spectral resolution of 4000, working from 1.51-1.82μm, and the second grating is for J-band with a resolution of 8000, from 1.19μm and 1.37μm. Compared with the VPH gratings that are used for optical spectrograph, these two VPH gratings have lower line frequencies of 177l/mm and 440l/mm for H-band and J-band, respectively. Kaiser Optical System Inc. (KOSI) and Wasatch Photonics (Wasatch) provided prototyping for each of these bands for a total of four gratings in our sample.

We present the performance characteristics of four near-infrared VPH gratings with specifications that are designed for the IRIS spectrograph. We will describe the experimental setup and method for measuring direct efficiencies at non-cryogenic temperatures. The relation between the efficiency versus angle range of VPH gratings is tested, and the entrance angular bandwidth governed by the Bragg angle is given. Our measurements are compared to theoretical efficiencies for these grating types. We also compare the VPH grating performance to reflective ruled diffraction gratings (Meyer et al, this conference) with similar properties to help in the selection of grating types and vendors for IRIS. This comparison will also be crucial for other current and future near-infrared spectrographs with a range of industrial applications.

## 2 VPH GRATING PROPERTIES

Different from surface relief gratings, VPH gratings are transmission gratings usually with a 10 to 100 micron dichromated gelatin (DCG) layer placed between two glass substrates. The chosen glass substrate combination is dependent on the desired wavelength of light. This sandwich structure makes VPH gratings more stable and durable than surface relief gratings, since the diffracting surface is protected by glass, which can be easily cleaned. The DCGs are generated through exposure to a laser interferogram, the interferometric pattern is recorded within the volume of gelatin material, so the refractive index is modulated in a sinusoidal fringe pattern[8]. When light passes through the VPH grating, the refractive index modulation generates optical path differences for different wavelengths, and the diffraction that occurs follows the grating equation. The particular wavelength of light that satisfies Bragg's law has the highest efficiency, which is dependent on the incidence angle. The VPH gratings have lower scattered light than ruled gratings since the diffraction is a function of the modulation of the gelatin material, rather than the surface quality of the ruled grating and individual facets. This means that VPH gratings can have very high efficiencies designed for particular incidence angles and wavelengths of light[20]. What is important, the high efficiency and large super-blaze (the envelope of the entire efficiency spectrum according to the incidence angle is called the super-blaze[22]) make VPH gratings attract much attention, especially in high resolution applications. Superblaze refers to the ability of VPH grating instrument to tune for each central wavelength setting. As VPH gratings are operated at Littow configuration, the spectrograph detector must also be rotated or articulated about the grating center. In this way, high efficiency and high resolution can be achieved over a broad band.

Transmission gratings can also bring great improvements and advantages for optical designs. It makes the optical structure flexible and compact, and easy to work with for larger incidence angles system (with combined prism). However, the efficiency will decrease much when the incidence angle deviates the Bragg angle. Depending on current manufacture process, if the VPH gratings can achieve their theoretical performance is what we concerned. In order to understand the relationship between the characters and its manufacture, the principle of VPH gratings is introduced as following.

The incidence light is diffracted due to the refractive index periodic modulation of the gelatin film material as shown in Figure 1. Therefore, the fundamental parameters that relate the efficiency are the thickness of the active film ($d$) and the modulation of the refractive index ($\Delta n_1$). The grating parameters include incidence angle $\alpha$, diffraction angle $\beta$, slant angle $\gamma$, grating thickness $d$, the grating period $\Lambda$, the frequency of the intersection of the grating $\nu$, grating vector $K$, and index of refraction $n$.

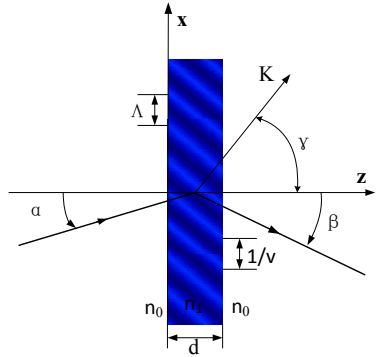

Figure 1: VPH grating geometry and parameters with the DCG (include incidence angle $\alpha$, diffraction angle $\beta$, slant angle $\gamma$, grating thickness $d$, the grating period $\Lambda$, the frequency of the intersection of the grating $\nu$, grating vector $K$, and index of refraction $n$), the glass substrates on either side are not shown.

The diffraction follows the classical grating equation:

$$m\lambda = \Lambda(\sin\alpha_0 + \sin\beta_0) \qquad (1)$$

Where $m$ is the diffraction order, and $\lambda$ is the wavelength.

If the fringes are not slanted, thus $\gamma = \pi/2$, $\Lambda = 1/\nu$. The Bragg condition is met when the incidence angle and diffraction angle are equal and opposite. The Bragg equation can be given by:

$$m\lambda = 2\Lambda n_1 \sin\alpha_1 \qquad (2)$$

where

$$n_1 \sin\alpha_1 = n_0 \sin\alpha \qquad (3)$$

From the Bragg's law and the Kogelnik's two-wave coupled-wave theory, most VPH grating properties can be deduced[23]. According the Kogelnik's two-wave coupled-wave theory, the most important conclusion is that the diffraction efficiency $\eta$, spectral bandwidth $\Delta\lambda$, and angular bandwidth $\Delta\alpha$ are a function of the grating modulation $\Delta n_1$ and actual film thickness[23,24]. This brief conclusion has straight effect on VPH grating manufacture process. For a plane transmission grating without slanted fringes, the TE mode (polarization perpendicular to the incidence plane) diffraction efficiency at the first Bragg order is given by *Kogelnik* 1969:

$$\eta_{TE} = \sin^2[\pi\Delta n_1 d/(\lambda\cos\alpha_1)] \qquad (4)$$

For the TM mode (polarization parallel to the plane of incidence), the diffraction efficiency equation becomes:

$$\eta_{TM} = \sin^2[\pi\Delta n_1 d\cos(2\alpha_1)/(\lambda\cos\alpha_1)] \qquad (5)$$

The polarization efficiency dependency on the VPH gratings occurs through the reduced effective coupling constant $\cos(2\alpha_1)$. If the incidence angle $\alpha_1$ in the material is small, the polarization dependencies are minimized.

The first order spectral bandwidth and angular bandwidth are approximated by:

$$\Delta\lambda_{FWHM}/\lambda \propto (\Lambda/d)\cot\alpha_1 \qquad (6)$$
$$\Delta\alpha_{FWHM} \propto \Lambda/d \qquad (7)$$

Where $\Delta\alpha_{FWHM}$ is expressed in radians.

From the above relations, we can see that in order to achieve the proper efficiency and bandwidth, the right index modulation ($\Delta n_1$) and the precise thickness ($d$) of VPH grating are necessary. Having a highly dispersive grating is fine, but increasing the line density has an influence on the bandwidth. To maintain an acceptable bandwidth, when $\Lambda$ is reducing, we must reduce the grating thickness *d* at the same time to guarantee the appropriate bandwidth. Besides, the coupled-wave equations, the grating thickness including the substrates must be accounted for the total efficiency. Therefore, to keep the efficiency at highest value, when the grating thickness is reduced, the index modulation needs to be increased.

In the infrared wavelength, the high index modulation is important. From equation (4) and (5), if the wavelength increases, the refractive index modulation must increase to keep the diffraction close to unity. For infrared wavelengths to satisfy the Bragg's law with a specific incidence angle, VPH gratings need to have lower groove densities. The manufacturing of these low groove density gratings are challenging since the equipment needs a small opening-angle and large aperture mirrors that need low wavefront error. Infrared instruments usually work in low temperatures, so the cryogenic behavior of the VPH grating is very significant. Form Tamura et al. 2006, they completed 5 cycles between room temperature and 200K and they found that the diffraction efficiency and angular dispersion are nearly independent of temperature[25].

## 3 VPH GRATING MEASUREMENT

Based on the current optical design, the spectrograph camera focal length is 370mm, and the pixel pitch of the detector (Teledyne Hawaii-4RG) is 15μm. The opening angle of spectrograph is 45°, and the spectral resolution has three levels, R=4000, R=8000 and R=10000 which are specified for 2 pixels. According the requirements of IRIS, the main parameters of grating for each bandpass were calculated[6]. We select H-band R=4000 and J-band R=8000 to investigate the efficiency and performance of our gratings. The VPH gratings for H-band R=4000 and J-band R=8000 have relatively lower groove densities, and are therefore more difficult to manufacture. The specifications are in Table 1. Kaiser Optical System Inc. and Wasatch Photonics provided prototyping for each of these gratings. According the IRIS requirements for the VPH gratings, KOSI and Wasatch designed the VPH gratings with specific parameters and give the theoretical simulation by Rigorous Coupled-Wave Analysis (RCWA). Both of their simulation results are similar. The peak efficiency of H-band is round 85% at 1.629μm, with 25% decrease when 2.5 °deviation from the Bragg angle. For J-band, the peak efficiency of 1.27μm is 95%, with 25% drop in efficiency when 5° off the Bragg angle.

Our team measured the efficiencies of these gratings at a single wavelength per passband. The peak efficiency at different incidence angles and efficiency distribution at each order at both the Bragg angle and off-Bragg angle were measured. At the end, we compare the VPH gratings performance between the theoretical simulation and actual measurement results.

Table 1. Specifications of VPH gratings designed for IRIS

|  | IRIS H-band R=4000 | IRIS J-band R=8000 |
| --- | --- | --- |
| **Central Wavelength** | 1.629 μm | 1.27 μm |
| **Bandpass** | 1.51-1.82 μm | 1.19-1.37 μm |
| **Fringe Frequency** | 177 l/mm | 440 l/mm |
| **Incidence Angle** | 8.3° (at 1.629μm) | 16.2°(at 1.27μm) |
| **Grating Clear Aperture** | 100mm×100mm | 100mm×100mm |

### 3.1 Measurement Equipment and Stability

For the infrared gratings efficiency measurement, we used an InGaAs camera (Raptor Photonics OWL SW 1.7 CL320) as the detector and 1.55μm and 1.31μm laser diodes as the light source for H-band and J-band, respectively. In order to guarantee the stability of laser source, we use LM9LP-LD/TEC as the laser driver. The stability of each laser diode was measured for six hours continuously. The laser diode needs to be turned for ~1 hour before it stabilizes in its power. After the first hour of continuous operation, the standard deviation of the measured intensity is within 0.3%.

The laser diode is connected to a collimator (F240FC) through a SMF-28 fiber. SMF-28 is a single mode fiber, and its core diameter is 8.2μm, operating from 1.310μm to 1.625μm. We used a variable attenuator (VOA50-FC-SM) to adjust the intensity of the incidence light and make sure the final spot on the detector is unsaturated. Since there is big difference between the brightest order and the faintest orders, the exposure time is varied from each order to achieve a good signal-to-noise ratio. When the polarized efficiency was measured, the linear polarizer (LPNIR050-MP) is installed before the VPH grating. After the VPH grating, an achromatic lens focuses the beam on the detector that sits on an optical rail system. The diameter of the laser without the grating is ~1.5mm with little angle deviation. An anodized aluminum baffle box resides over the entire experiment to eliminate scattered background light. The schematic of configuration is shown in Figure 2.

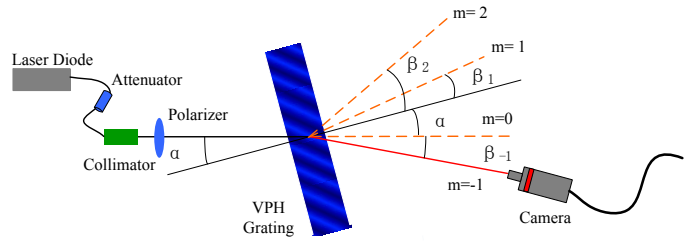

Figure 2. Schematic of the configuration. The laser diode is coupled to the fiber, and the total intensity is adjusted by an attenuator. We collimate the laser beam before it passes through the (polarized and) VPH grating. Adjust the position and angle of camera, and make sure the diffracted orders focus on the center of the detector.

**3.2 Adjustment Process for Measurement**

The peak efficiency for a given order is defined as the measured flux of a monochromatic light source diffracted into that order relative to the total incidence flux. The typical efficiency measurement procedure is as follows: (1) Set up the VPH grating perpendicular to the incidence light beam; (2) rotate the VPH grating to the specific incidence angle; (3) adjust the camera position and angle to measure the diffracted order; (4) change the gain of the attenuator to get a reasonable incidence intensity; (5) take 100 frames of the flux with the best exposure time; (6) block the laser, and take the background of the setup with the same exposure time; (7) repeat the above steps for each order; (8) remove the VPH grating, and measure the pure incidence flux of the laser; and (9) using the flux of the specific diffracted order to divide the total incidence flux of the laser, the efficiency will be calculated. It is important to note that we observed reflected light off the incoming light surface of the VPH grating, which would decrease the total efficiency.

The whole process takes roughly one hour. Within this time, the polarization state of fiber does not change and the temperature of the camera does not increase[26]. Because VPH grating efficiencies are quite sensitive to the incidence angle, the precision of the incidence angle is very important for our measurement. In our setup, the incidence angle precision is less than 10′. Due to reflective and scattered light, the sum of the efficiencies of all diffracted orders is less than 100%. We confirm this result by the following data in Table 2 to 5.

**3.3 Efficiency Results**

We have four VPH gratings, two at H-band and two at J-band. The clear aperture is 100mm×100mm. Herein, we only provide the efficiency results of the central area. In order to evaluate the performance of the whole grating, we will measure many other locations of the grating. We measured the peak efficiency of different incidence angles, and find the actual Bragg angle of each VPH grating, and then measured each order's efficiency at the Bragg angle.

**3.3.1 H-band VPH grating Efficiency measurement**

We used a 1.55μm laser diode to measure the efficiency at different incidence angles for the TM mode, TE mode, and without the polarizer. The results of the KOSI H-band VPH grating are shown in Figure 3 and 4, and the Wasatch H-band VPH grating are shown in Figure 5 and 6.

From the efficiency curves in Figure 5, we can find that the KOSI H-band VPH grating is not sensitive to polarization, and there is little deviation between the TM and TE polarization states. Its peak efficiency at the Bragg angle (~7.5 °) is 88%. The efficiency drops 20%-23% when the incidence angle is 2.5° off the Bragg angle. We also measure the efficiency of each order at the Bragg angle 7.5°, and the details of these results are listed in Table 2. The same measurement was done at off Bragg angle (8°, 9°) too, and the comparison is in Figure 4. The peak efficiencies in Figure 3 and 4 come form different measurement, so there is some deviation between them, which is within the measurement error range.

In Figure 5, it presents the peak efficiency at different incidence angles of Wasatch H-band VPH grating. Its peak efficiency at the Bragg angle (~10 °) is between 61% and 66%. From the efficiency curve, we can see that the efficiency drops more than 35% when the incidence angle is 5° off Bragg angle. The efficiency of each order at the Bragg angle 10 ° and off Bragg angle (9°, 11 °) is shown in Figure 6. The details of the Wasatch H-band efficiencies at the Bragg angle are shown in Table 3.

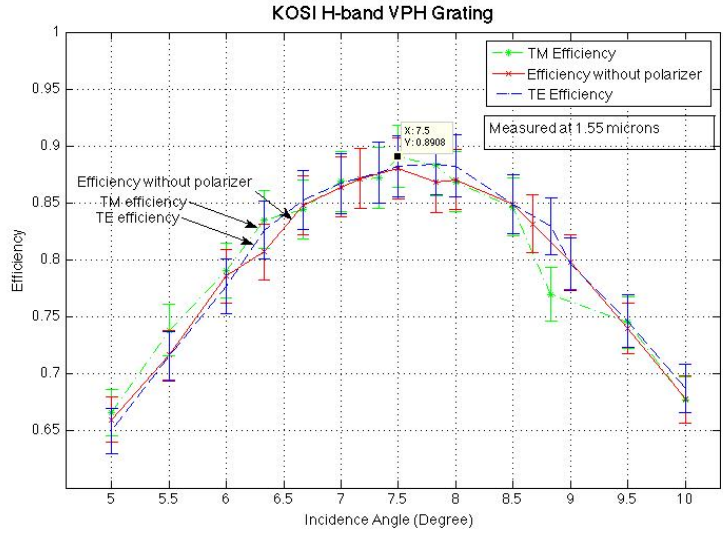

Figure 3. KOSI H-band efficiency at different incidence angles. The peak efficiency of 1.55μm is around 88% at the Bragg angle (7.5°) with the error 2.7%. The efficiency drop 20-23% when the incidence angle is 2.5° off from the Bragg angle in different modes.

Table 2 KOSI H-band efficiency of each order at the Bragg angle (7.5°)

| Order | 3 | 2 | 1 | 0 | -1 | -2 |
|---|---|---|---|---|---|---|
| **TM Efficiency (%)** | 0.85±0.03 | 1.51±0.05 | 87.50±2.66 | 5.24±0.16 | 2.71±0.08 | 0.33±0.01 |
| **UNP Efficiency (%)** | 1.27±0.04 | 1.16±0.03 | 88.88±2.72 | 3.73±0.11 | 1.93±0.06 | 1.03±0.03 |
| **TE Efficiency (%)** | 1.59±0.04 | 0.79±0.02 | 87.27±2.65 | 2.34±0.07 | 1.24±0.04 | 1.67±0.05 |

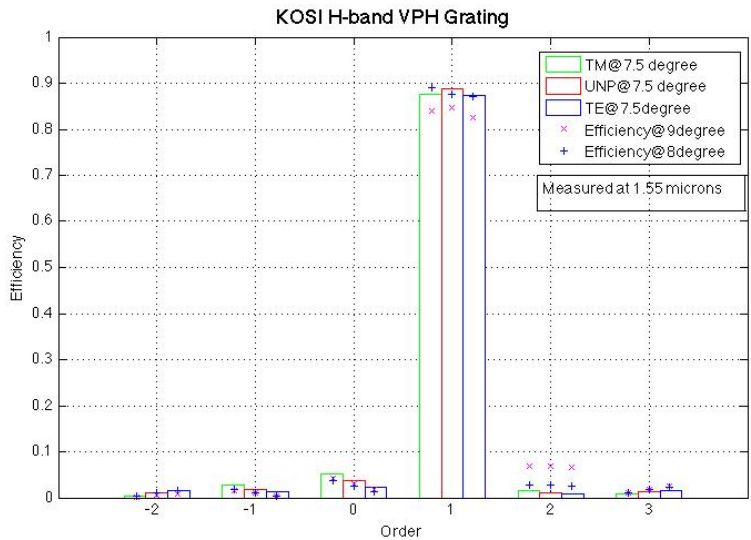

Figure 4. KOSI H-band efficiency comparison at Bragg angle (7.5 °) and off Bragg angles (8° blue, 9° magenta)

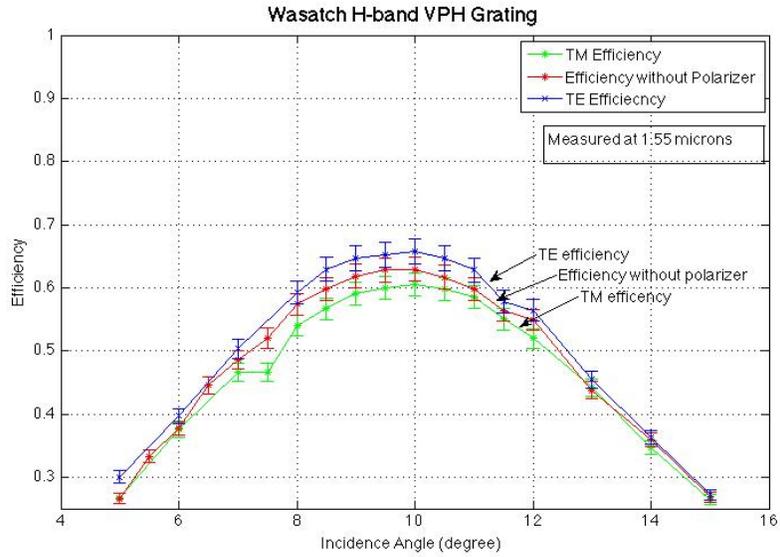

Figure 5. Wasatch H-band efficiency at different incidence angles. The peak efficiency of 1.55μm is between ~61%-66% at the Bragg angle (10°) with the error ~1.9%. The efficiency drop more than 35% when the incidence angle is 5° off from the Bragg angle.

Table 3 Wasatch H-band efficiency of each order at Bragg angle (10°)

| Order | 3 | 2 | 1 | 0 | -1 | -2 |
|---|---|---|---|---|---|---|
| **TM Efficiency (%)** | 2.22±0.07 | 7.61±0.23 | 61.33±1.86 | 6.79±0.20 | 12.66±0.38 | 3.32±0.10 |
| **UNP Efficiency (%)** | 2.73±0.08 | 6.59±0.20 | 63.35±1.93 | 6.17±0.19 | 10.68±0.33 | 4.16±0.13 |
| **TE Efficiency (%)** | 3.79±0.12 | 4.84±0.15 | 67.33±2.05 | 5.21±0.16 | 7.20±0.22 | 5.83±0.18 |

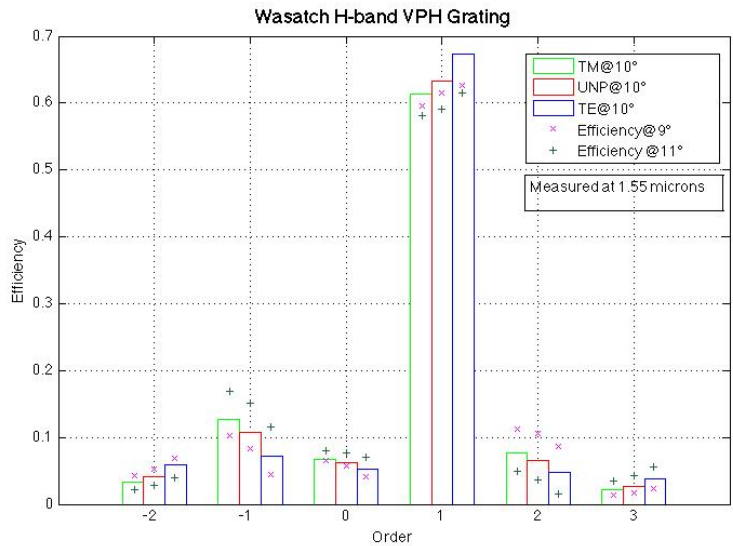

Figure 6. Wasatch H-band efficiency comparison at Bragg angle (10 °) and off Bragg angles (9° magenta, 11° blue)

### 3.3.2 J-band VPH grating Efficiency measurement

We used a 1.31μm laser diode to measure the efficiency at different incidence angles for the TM mode, TE mode, and without the polarizer. The results of the KOSI J-band VPH grating are shown in Figure 7 and 8, and the performance of

Wasatch J-band VPH grating is shown in Figure 9 and 10. The efficiency of each order at the Bragg angle was measured. There is 10% difference between each of the polarization modes for KOSI J-band VPH grating. According to the VPH grating principles, this deviation between polarization may be due to the effective of the coupling constant $\cos(2\alpha_1)$. If the incidence angle $\alpha_1$ within the medium is small, the polarization dependencies are minimized.

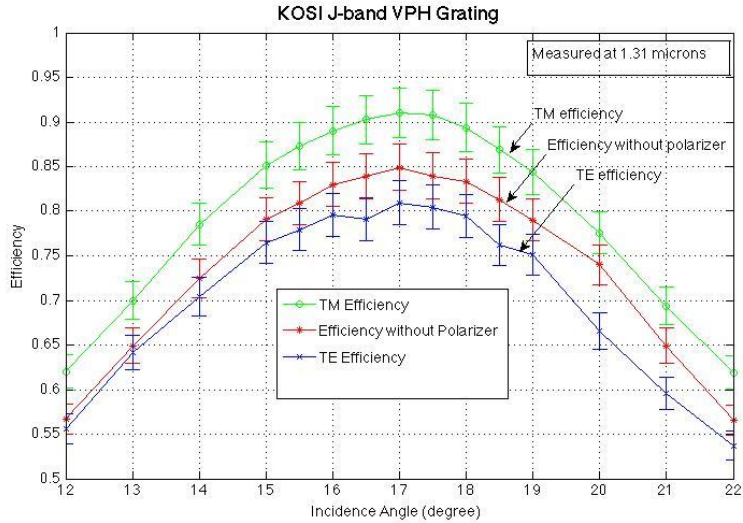

Figure 7. KOSI J-band efficiency at different incidence angles. The TM peak efficiency of 1.31μm is 91% at the Bragg angle (17°) with 2.8% error, while the TE peak efficiency is 81% with 2.4% error. The efficiency drop 25%-28% when the incidence angle is 5° off from the Bragg angle.

Table 4. KOSI J-band efficiency of each order at Bragg Angle (17°)

| Order | 2 | 1 | 0 | -1 |
|---|---|---|---|---|
| **TM Efficiency (%)** | 1.55±0.05 | 89.49±2.78 | 3.78±0.12 | 2.49±0.08 |
| **UNP Efficiency (%)** | 0.87±0.03 | 83.45±2.54 | 9.14±0.28 | 3.40±0.10 |
| **TE Efficiency (%)** | 0.81±0.03 | 78.92±2.39 | 10.19±0.31 | 3.47±0.11 |

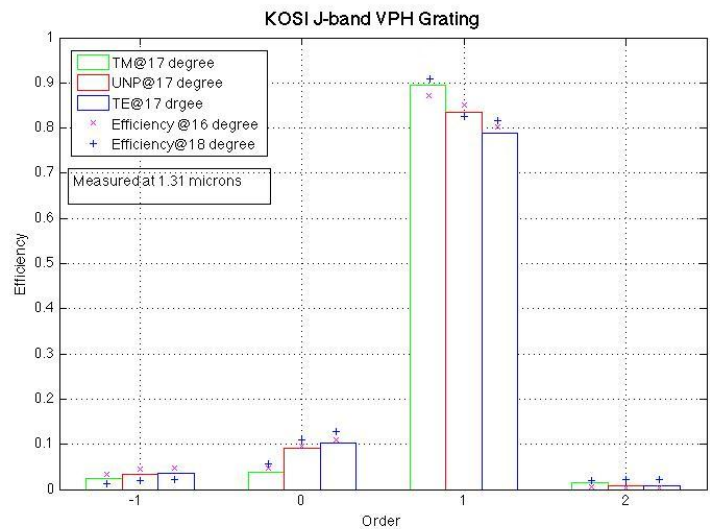

Figure 8. KOSI J-band efficiency comparison at Bragg angle (17°) and off Bragg angles (16° magenta, 18° blue)

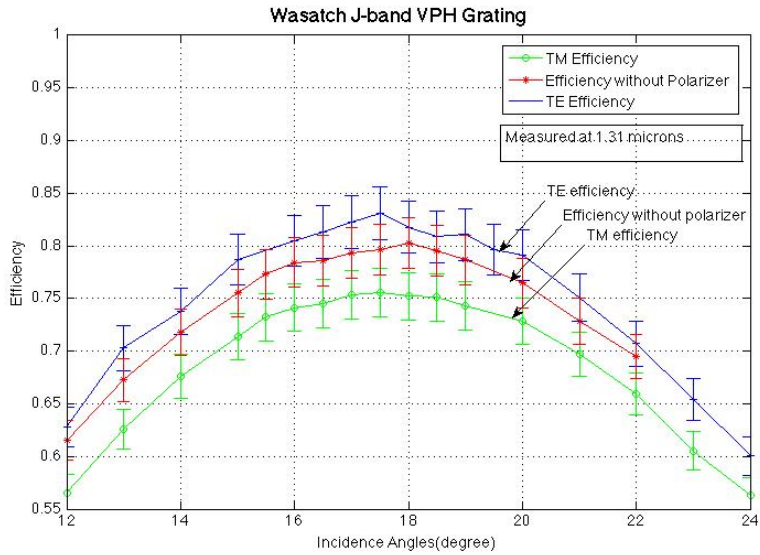

Figure 9. Wasatch J-band efficiency at different incidence angles. The TM peak efficiency of 1.31μm is ～75% at the Bragg angle (17.5°) with 2.3% error, while the TE peak efficiency is～82% with 2.5% error. The efficiency drop 13-16% when the incidence angle is 5° off from the Bragg angle.

Table 5. Wasatch J-band efficiency of each order at Bragg Angle (17.5°)

| Order | 2 | 1 | 0 | -1 |
|---|---|---|---|---|
| **TM Efficiency (%)** | 0.10±0.003 | 74.99±2.27 | 10.84±0.32 | 2.96±0.09 |
| **UNP Efficiency (%)** | 0.50±0.02 | 80.32±2.43 | 5.9±0.18 | 4.01±0.12 |
| **TE Efficiency (%)** | 0.60±0.02 | 80.84±2.45 | 3.95±0.12 | 4.15±0.13 |

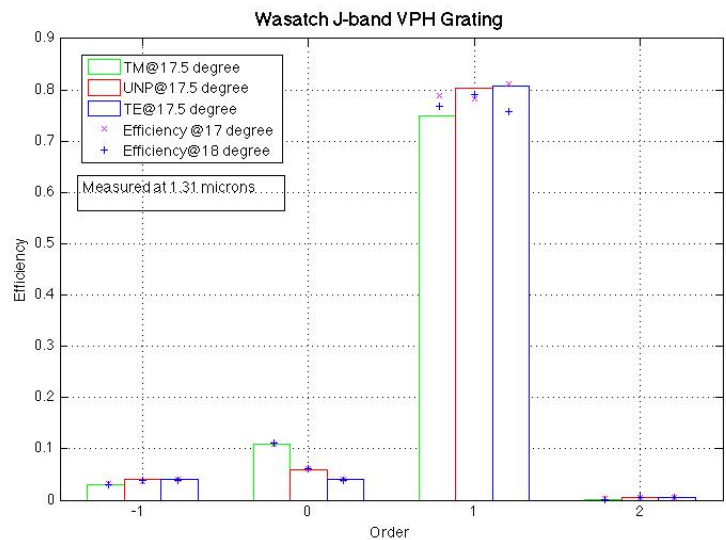

Figure 10. Wasatch J-band efficiency comparison at Bragg angle (17.5°) and off Bragg angles (17° magenta, 18° blue)

**3.4 Measurement Uncertainties and Error Analysis**

The dominating error terms are the stability of the laser and infrared camera during each of the measurements. The laser diode intensity stability is 0.3% and the fluctuation due to camera temperature changes is 0.3% *(measured by Mieda et al. 2014)*[26]. For our uncertainties, we also consider the flat field error, background deviation, dark current deviation, readout noise, random noise and Poisson noise. We also include a systematic error, from the accuracy of the incidence angle at the VPH grating is 10' and the camera position error is 2.0%. During the measurement, we try to make the setup background dim by using baffle box, and cooled the camera to minimize the dark current. Even though, we find the variance is not zero when there is no light entrance. The deviation of the mean values of the setup background is 2 DN (digital number). From the absolute difference between the average dark count and linear fit plotted as a function of exposure time, there are always 10 DN fluctuations regardless of exposure time *(the same result as that measured by Mieda et al. 2014)*[26]. We look 10 DN as the random noise of camera. Hence, the detector photoelectron noise is the function of the Poisson noise of the signal, the setup background, the dark current, and readout noise, and camera random noise in the spot aperture. Using error propagation, we derive the error for each efficiency measurement. We also measured the VPH grating at least three times, so the final reported measurements of the peak efficiency in Table 6 and 7 was the weighted average.

## 4 COMPARISON WITH REFLECTIVE GRATING

Meyer et al, this conference, presents the reflective ruled diffraction gratings with comparable properties. The main properties of H-band gratings are listed in Table 6. The final peak efficiencies of VPH gratings are the averages of three times results. From the efficiency measurements, KOSI H-band VPH grating has the highest efficiency of ~88% both in TM and TE modes, with little polarization effect, but it drops by 20-23% at 2.5° off the Bragg angle. CIOMP (Changchun Institute of Optics, Fine Mechanics, and Physics, Chinese Academy of Sciences) H-band ruled grating also have high efficiencies, but show larger differences of ~15% between TM and TE modes. The same comparison is done for J-band R8000. Only Bach provided ruled grating for this band. KOSI J-band VPH grating has highest efficiency with the TM efficiency of 90.23%. Wasatch J-band VPH grating has lower efficiency, but better performance in polarization (6.7% deviation between TE and TM modes) and bandpass (less efficiency drop than KOSI J-band VPH grating). Bach J-band ruled grating have the lower efficiencies compared VPH gratings. From these two tables, we can find the VPH gratings of KOSI have higher efficiencies than those of Wasatch, while with much more drops in efficiencies at both H-band and J-band.

Table 6. H-band R4000 Gratings Performance Comparison. The final efficiencies and error bars of VPH gratings are the averages of three measurements, and the final results of Ruled gratings are the single measurement.

| Grating | Peak Efficiency (TM and TE) | Efficiency drop from Bragg/Blazed Angle | Clear Aperture |
|---|---|---|---|
| Theoretical Performance Simulated by RCWA | 85% 85% | 20% at 2.5° from Bragg angle, 60% at 5° from Bragg angle | |
| KOSI H-band VPH grating | 88.29±0.89% 87.75±0.88% | 20-23% at 2.5° from Bragg angle | 100mm× 100mm |
| Wasatch H-band VPH grating | 60.90±0.61% 66.51+0.67% | 35-40% at 5° from Bragg angle | 100mm× 100mm |
| Bach H-band Ruled grating | 70.78±2.68% 71.25±3.07% | | 25mm× 25mm |
| CIOMP H-band Ruled grating | 98.90±3.36% 80.71±2.62% | Within 3% | 50mm× 50mm |

Table 7. J-band R8000 Gratings Performance Comparison

| Grating | Peak Efficiency (TM and TE) | Efficiency drop from Bragg/Blazed Angle | Clear Aperture |
|---|---|---|---|
| Theoretical Performance Simulated by RCWA | 95% 95% | 25% at 5° from Bragg angle | |
| KOSI J-band VPH grating | 90.23±0.99% 79.91±0.83% | 25-28% at 5° from Bragg angle | 100mm× 100mm |
| Wasatch J-band VPH grating | 75.26±0.79% 81.96±0.98% | 13-16% at 5° from Bragg angle | 100mm× 100mm |
| Bach J-band Ruled grating | 75.18±2.42% 78.78±2.54% | | 25mm×25mm |

## 5 CONCLUSION

From the principle analysis, VPH gratings can achieve extremely high efficiencies (>90%) and has indicated by theoretical predictions are not sensitive to polarization. But the efficiencies decrease much when the incidence angle is off the Bragg angle. We measured four near-infrared VPH gratings with the properties designed for IRIS, and we find the peak efficiencies of these two types of gratings are in good agreement to theoretical prediction. The best peak efficiency of 1.55μm at the Bragg angle is ~88±0.9% in both TM and TE modes, without apparent polarization effect at H-band, which have obvious advantages compared to ruled gratings. The efficiency drops 20-23% when 2.5° deviation from the Bragg angle. The peak efficiency and the decrease in efficiency are correlated with the theoretical simulation. For J-band, the highest peak efficiency of 1.31μm at the Bragg angle is 90.23±0.99% at TM mode and 79.91±0.83% at TE mode, much higher than ruled grating, which is larger than the RCWA prediction. The drop in efficiency is similar as the prediction, ~25% decrease when 5° off the Bragg angle. The VPH gratings of KOSI have higher peak efficiencies than those of Wasatch, but Wasatch VPH gratings have better performance at larger deviation off Bragg angel.

From our testing, we find the transmission of VPH gratings is substantially higher than ruled gratings. However, the efficiencies of VPH gratings drop dramatically when the incidence angle is off the Bragg angle. In the IRIS design, there is a varied angle of incidence in the spectral direction on the grating. For the lenslet, the incidence angle is up to $8.78°^{6}$, and the angles off axis will be up to ±4.4° off the Bragg angle. Therefore, this is a concern for VPH grating performance for the IRIS design. From the grating manufacture, VPH gratings are more feasible to achieve extra-large apertures compared with ruled gratings. All of these factors are what IRIS will consider for the final design.

It is important to note that our measurements are made with a monochromatic light source and does not cover the full bandpass, and therefore the performance of the grating is limited. VPH gratings are sensitive to the incidence angle, thus when the angle changes the efficiency of a particular wavelength drops but another wavelength efficiency (which yield to the Bragg condition at the new incidence angle) will increase. It means VPH gratings can diffract a large range of wavelengths (Δλ), but in very specific direction (small Δα). The super-blaze curve will provide much more efficiency information for high spectral resolution applications.

## ACKNOWLEDGEMENTS
We want to thank Kaiser Optical System Inc. and Wasatch Photonics for working with us on procuring these gratings. Each company was incredibly helpful and responsive in the design and procurement of these challenging gratings. During the measurement, they gave us suggestions and feedback, which guaranteed the projects success. We thank the support and resources offered by the Dunlap Institute of Astronomy & Astrophysics at University of Toronto. Shaojie Chen is supported through a Dunlap Fellowship from the Dunlap Institute for Astronomy & Astrophysics, University of Toronto. The Dunlap Institute is funded through and endowment established by the David Dunlap family and the University of Toronto. The authors gratefully acknowledge the support of the TMT partner institutions. They are the Association of Canadian Universities for Research in Astronomy, California Institute of Technology, Department of Science and Technology India, National Astronomical Observatories of the Chinese Academy of Science, the National Astronomical Observatory of Japan, and the University of California. The TMT project is planning to build the telescope facilities on Mauna Kea, Hawaii. The authors wish to recognize the significant cultural role and reverence that the summit of Mauna Kea has always had with the indigenous Hawaiian community.